\documentclass[aps,pra,groupedaddress,showpacs,showkeys]{revtex4}

\usepackage{amsfonts}
\usepackage{amsmath}
\usepackage{amssymb}
\usepackage{graphicx}
\usepackage{epsfig}

\newtheorem{theor}{Theorem}

\allowdisplaybreaks

\begin{document}


\title{Lindbladian Evolution with Selfadjoint Lindblad Operators as Averaged Random Unitary Evolution}


\author{D. Salgado}
\email[]{david.salgado@uam.es}
\affiliation{Dpto. F\'{\i}sica Te\'{o}rica, Universidad Aut\'{o}noma de Madrid\\
28049 Cantoblanco, Madrid (Spain)}

\author{J.L. S\'{a}nchez-G\'{o}mez}
\email[]{jl.sanchezgomez@uam.es}
\altaffiliation{Permanent Address}
\affiliation{Dpto. F\'{\i}sica Te\'{o}rica, Universidad Aut\'{o}noma de Madrid\\
28049 Cantoblanco, Madrid (Spain)}


\date{\today}

\begin{abstract}
It is shown how any Lindbladian evolution with selfadjoint Lindblad operators, either Markovian or nonMarkovian, can be understood as an averaged random unitary evolution. Both mathematical and physical consequences are analyzed. First a simple and fast method to solve this kind of master equations is suggested and particularly illustrated with the phase-damped master equation for the multiphoton resonant Jaynes-Cummings model in the rotating-wave approximation. A generalization to some intrinsic decoherence models present in the literature is included. Under the same philosophy a proposal to generalize the Jaynes-Cummings model is suggested whose predictions are in accordance with experimental results in cavity QED and in ion traps. A comparison with stochastic dynamical collapse models is also included. 
\end{abstract}


\maketitle



\section{Introduction}

Since the early years of quantum mechanics \cite{Dirac58a} the principle of
quantum superposition has been recognized to play a prominent role in the
theory and its applications. The destruction and preservation of these
superpositions of quantum states occupy a central place in issues such as
the quantum-to-classical transition \cite{Zur91a,GiuJooKieKupStaZeh96a} and potential
technological applications in Quantum Information, Computation and
Cryptography \cite{BouEkeZei00a,LoPopSpi98a,ChuNie00a}. From a physical
standpoint the loss of coherence in quantum systems is rooted on the
pervasive action of the environment upon the system. This environmental
action has received a careful mathematical treatment (cf. \cite{Dav76a,AliLen87a,BrePet02a} and multiple references therein)
going from a constructive approach based on disregarding the degrees of
freedom of the environment due to their lack of control by the
experimenter ("tracing-out" methods) to an axiomatic approach based on the
initial setting of physically motivated axioms to derive an appropiate
evolution (master) equation for the system \cite{Lin76a,GorKosSud76a}.\\
Most of these master equations (ME's hereafter) satisfy the Markov
approximation (semigroup condition) and can be put into the Lindblad form
:

\begin{equation}
\frac{d\rho}{dt}=-i[H,\rho(t)]+\frac{1}{2}\sum_{j}\left\{[V_{j}\rho(t),V_{j}^{\dagger}]+[V_{j},\rho(t)V_{j}^{\dagger}\right\}
\end{equation}

\noindent where $H$ is the Hamiltonian of the system and $\{V_{j}\}$
are operators (so-called Lindblad operators) containing the effect of the
environment upon the system. Indeed in the axiomatic approach the Markov
approximation is posed as an initial hypothesis \cite{Lin76a}, thus
rendering highly difficult a generalization to nonMarkovian situations.\\

In this work we develop a novel attempt to derive ME's both in the markovian
and the nonMarkovian regimes using stochastic methods
\cite{Oksendal98a,KarShr91a} jointly with
well-known operator techniques commonly used in quantum mechanics
\cite{Lou73a}. The main idea consists of building \emph{random}
evolution operators (evolution operators with one or several stochastic
parameters in it) which contains the decohering effect of the environment
and then taking the stochastic expectaction value with respect to this
(uncontrollable) randomness. The paper is organized as follows. In section
\ref{LindAsRand} we state and prove our main (though still somewhat partial) result,
namely that any Lindblad-type ME, either Markovian or nonMarkovian, with
selfadjoint Lindblad operators can be understood as an averaged random
unitary evolution. In section \ref{AnalCons} we discuss some first mathematical
consequences of this result such as a very fast method to solve ME's
provided the unitary solutions are known; we illustrate this by solving
the phase-damping ME for the multiphoton resonant Jaynes-Cummings model in the rotating-wave approximation \cite{Sin82a}
(section \ref{SolDampJCM}). We then comment in section \ref{NonMarkEvol} two immediate
consequences, namely both Markovian and nonMarkovian regimes are
attainable under the same mathematical formalism and the Lindbladian
structure with selfadjoint Lindblad operators is shown to have an origin
independent of the Markov approximation. In section \ref{IntrDecoh} we show how the
flexibility of the mathematical language employed can easily generalize
some intrinsic decoherence models present in the literature
\cite{Mil91a,BonOliTomVit00a}. In section \ref{RabiQED} we discuss the previous main
result under a more physical spirit by proposing a slight generalization
of the Jaynes-Cummings model (section \ref{StoJCM}), comparing this proposal with
experimental results in optical cavities experiments (section \ref{QED}) and
finally (section \ref{IonDecay}) showing how the proposed formalim can account for
reported exponential decays of Rabi oscillations in ion traps. We include
in section \ref{Discuss} some important comments regarding a brief comparison with
existing models of stochastic evolution in Hilbert space, the possibility
of intrinsic decoherence phenomena and future prospects. Conclusions and a short appendix close the paper.

\section{Lindblad Evolution as an Averaged Random Unitary Evolution}
\label{LindAsRand}
The main result whose consequences are to be discussed below is the following: \emph{Every Lindblad evolution with selfadjoint Lindblad operators can be understood as an averaged random unitary evolution}. We will analyse this proposition in detail. The objective is to reproduce the Lindblad equation\footnote{From the beginning it will  be taken into account that $V_{i}^{\dagger}=V_{i}\quad  i=1,\dots,n$.} 

\begin{eqnarray}
\frac{d\rho(t)}{dt}&=&-i[H,\rho(t)]+\frac{1}{2}\sum_{i=1}^{n}\{[V_{i}\rho(t),V_{i}]+[V_{i},\rho(t)V_{i}]\}\nonumber\\
\label{LindEq}&=&-i[H,\rho(t)]-\frac{1}{2}\sum_{i=1}^{n}[V_{i},[V_{i},\rho(t)]]
\end{eqnarray}

\noindent by adequately modifying chosen parameters in the original evolution operator. For simplicity let us start by considering the case $n=1$. We will first study the case where the Hamiltonian $H$ and the (selfadjoint) Lindblad operator $V$ commute. It is very convenient to introduce the following notation. The commutator between an operator $G$ and $X$ will be denoted by $\mathcal{C}_{G}[X]\equiv[G,X]$. Thus the von Neumann-Liouville operator will be $\mathcal{L}=-iC_{H}$, where $H$ denotes the Hamiltonian ($\hbar=1$). Then the Lindblad equation \eqref{LindEq} with $n=1$ can be arrived at by

\begin{enumerate}
\item Adding a stochastic term $\mathcal{B}_{t}V$ to the argument of the evolution operator:

\begin{equation}\label{AddTerm}
U(t)=\exp(-itH)\to U_{st}(t)=\exp(-itH-i\mathcal{B}_{t}V)
\end{equation}

where $\mathcal{B}_{t}$ denotes standard real Brownian motion \cite{Oksendal98a}.
\item Taking the stochastic average with respect to $\mathcal{B}_{t}$ in the density operator deduced from $U_{st}(t)$:

\begin{equation}
\rho(t)=\mathbb{E}[U_{st}(t)\rho(0)U_{st}^{\dagger}(t)]
\end{equation}

where $\mathbb{E}$ denotes the expectation value with respect to the probability measure of $\mathcal{B}_{t}$.

\end{enumerate}  

The proof of this result is nearly immediate. Taking advantage of the commutativity of $H$ and $V$ and making use of theorem 3 in \cite{Lou73a} (cf. appendix; relation \eqref{TheoLouisell}) we may write for the density operator:

\begin{equation}
\rho(t)=\exp(t\mathcal{L})\mathbb{E}[\exp(-i\mathcal{B}_{t}\mathcal{C}_{V})][\rho(0)]
\end{equation}

Thus all we have to do is to calculate the expectation value of the random superoperator $\exp(-i\mathcal{B}_{t}\mathcal{C}_{V})$. Developing the exponential into a power series and recalling \cite{Oksendal98a} $\mathbb{E}[\mathcal{B}^{n}_{t}]=\frac{(2n)!}{2^{n}n!}t^{n}$ if $n$ is even and $\mathbb{E}[\mathcal{B}^{n}_{t}]=0$ otherwise, we arrive at

\begin{eqnarray}
\rho(t)&=&\exp(-t\mathcal{L})\exp(-\frac{t}{2}\mathcal{C}_{V}^{2})[\rho(0)]\nonumber\\
&=&\exp(-t\mathcal{L}-\frac{t}{2}\mathcal{C}_{V}^{2})[\rho(0)]
\end{eqnarray}

\noindent which produces the desired master equation:

\begin{equation}
\frac{d\rho(t)}{dt}=-i[H,\rho(t)]-\frac{1}{2}[V,[V,\rho(t)]]
\end{equation}

When the Hamiltonian $H$ and the Lindblad operator $V$ do not commute, the previous method is not suitable, since the calculation of the expectation value cannot be performed in the same way. A way to circumvent this problem is to proceed in the same way as before but in the Heisenberg picture. Thus let $U(t)=\exp(-itH)$ be the original evolution operator in the Schr\"{o}dinger picture. The corresponding evolution operator in the Heisenberg picture will trivially be $U_{H}(t)=I$. As before we proceed in steps:

\begin{enumerate}
\item\label{NonC1} Add a stochastic term to the argument of the evolution operator:
\begin{eqnarray}
U_{H}(t)&=&\exp(-it0)\to U_{H,st}(t)=\nonumber\\\label{AddTermNon}
&=&\mathcal{T}\exp(-i\int_{0}^{t}V_{H}(s)d\mathcal{B}_{s})
\end{eqnarray}

\noindent where $\mathcal{T}$ denotes time-ordering and $V_{H}(t)$ is the operator $V$ in the Heisenberg representation.

\item\label{NonC2} Take the stochastic average with respect to $\mathcal{B}_{t}$ in the corresponding density operator:

\begin{equation}
\rho_{H}(t)=\mathbb{E}[U_{H,st}(t)\rho(0)U_{H,st}^{\dagger}]
\end{equation}

\item\label{NonC3} Finally to arrive at \eqref{LindEq} change to the Schr\"{o}dinger representation.
\end{enumerate} 

A comment should be made. The stochastic term added to the original evolution operator in \eqref{AddTermNon} is a natural generalization of the one added in \eqref{AddTerm}. The Ito integration now appears as a consequence of the time dependence of the operator to be added: $V_{H}(t)$. Note that when $[H,V]=0$, $V_{H}(t)=V$ and the stochastic term reduces to $V\mathcal{B}_{t}$ as before.\\  
Now to perform the previous tasks is a bit more involved. Firstly combining relation \eqref{TheoLouisell} and the $\mathcal{T}-$operation, the step \ref{NonC1} can be carried over:

\begin{eqnarray}
\rho_{H}(t)&=&\mathbb{E}[\mathcal{T}e^{-i\int_{0}^{t}V_{H}(s)d\mathcal{B}_{s}}\rho(0)\mathcal{\bar{T}}e^{i\int_{0}^{t}V_{H}(s)d\mathcal{B}_{s}}]=\nonumber\\
\label{ExpValNonC}&=&\mathcal{T}\mathbb{E}[e^{-i\int_{0}^{t}\mathcal{C}_{V_{H}(s)}d\mathcal{B}_{s}}][\rho(0)]
\end{eqnarray}

The expectation value \eqref{ExpValNonC} can be evaluated by resorting to functional techniques \cite{Kampen81a, FeynmanHibbs65a}. Recall that the characteristic functional of a stochastic process $\chi_{t}$ is defined as 

\begin{equation}
G_{\chi}[k(t)]=\mathbb{E}\left[\exp\left(i\int k(t)\chi_{t}dt\right)\right]
\end{equation}

\noindent where $k(t)$ is an arbitrary real-valued function. In particular, for white noise $\chi_{t}=\xi_{t}$ (cf. \cite{Kampen81a})

\begin{eqnarray}
G_{\xi}[k(t)]&=&\mathbb{E}[\exp\left(i\int k(s)d\mathcal{B}_{s}\right)]=\nonumber\\
&=&e^{-\frac{1}{2}\int_{0}^{t}k^{2}(s)ds}
\end{eqnarray}

\noindent where $d\mathcal{B}_{t}=\xi_{t}dt$, $\mathbb{E}\xi_{t}=0$ and $\mathbb{E}[\xi_{t}\xi_{s}]=\delta(t-s)$ have been used. From this it is then clear that \eqref{ExpValNonC} can be written as 

\begin{equation}\label{DensOpHeis}
\rho_{H}(t)=\mathcal{T}e^{-\frac{1}{2}\int_{0}^{t}\mathcal{C}^{2}_{V_{H}(s)}ds}[\rho(0)]
\end{equation}

Back to the Schr\"{o}dinger picture, the master equation derived from \eqref{DensOpHeis} is

\begin{equation}
\frac{d\rho(t)}{dt}=-i[H,\rho(t)]-\frac{1}{2}[V,[V,\rho(t)]]
\end{equation}

The generalization to many Lindblad operators is elementary: all we have to do is to use the n-dimensional standard real Brownian motion \cite{Oksendal98a} $\vec{\mathcal{B}}_{t}=(\mathcal{B}_{t}^{1},\cdots,\mathcal{B}_{t}^{n})$. The strategy is the same.

\section{Analytical Consequences}
\label{AnalCons}
The first consequences one can derive from the previous result are of analytical fashion. As an immediate aplication we will show how the Jaynes-Cummings model with phase damping in the rotating-wave approximation can be solved provided we know the solution to the original Jaynes-Cummings model. As a second consequence we will discuss how the previous result can be generalized to nonMarkovian situations, thus providing a common language for both Markovian and nonMarkovian evolutions. Finally it is shown how existing intrinsic decoherence models are naturally generalized using this formalism.

\subsection{Solution of the Resonant Multiphoton Jaynes-Cummings Model with Phase Damping}
\label{SolDampJCM}
The Jaynes-Cummings model (JCM hereafter) \cite{JayCum63a, ShoKni93a} shows an undoubtable relevance in the study of quantum systems in different fields such as Quantum Optics, Nuclear Magnetic Resonance or Particle Physics. It is an exactly solvable model which allows us to study specifically quantum properties of Nature such as electromagnetic field quantization or periodic collapses and revivals in atomic population. The JCM describes the evolution of a two-level quantum system (the atom) interacting with a mode of the electromagnetic field under certain approximations (rotating wave approximation, dipole approximation, etc. -- cf. \cite{JayCum63a, ShoKni93a} for details). Usually in normal experimental conditions this will be an idealization and the environment should be taken into account, the effect of which can be very appropiately treated introducing a phase-damping term \cite{Lou73a}. Thus the master equation for this system will read

\begin{equation}\label{Master}
\frac{d\rho(t)}{dt}=-i[H,\rho(t)]-\frac{\gamma}{2}[H,[H,\rho(t)]]
\end{equation}

\noindent where $H$ is the Hamiltonian of the system and $\gamma$ is a damping constant. Here we will show how \eqref{Master} can be very easily solved when $H$ is the resonant multiphoton JC Hamiltonian (cf. \cite{KuaCheCheGe97a} for an alternative approach), i.e. when

\begin{equation}
H=\omega a^{\dagger}a+\omega_{0}S_{z}+\lambda(S_{-}a^{\dagger m}+S_{+}a^{m})
\end{equation}

\noindent where $\omega$ denotes the frequency of the field mode, $\omega_{0}$ is the atomic transition frequency, $\lambda$ is the atom-field coupling constant, $a^{\dagger}$ and $a$ are the mode creation and annihilation operators respectively, $S_{z}$ is the atomic-inversion operator and $S_{\pm}$ are the atomic ``raising'' and ``lowering'' operators. An exact resonance is assumed, thus $\omega_{0}=m\omega$.\\
We will focus in two quantities of relevant physical meaning, namely the atomic inversion $W(t)=\textrm{Tr}[\rho(t)S_{z}]$ and the photon number distribution at time $t$: $P_{n}(t)=\langle n|Tr_{A}\rho(t)|n\rangle$. To compare with methods found in the literature \cite{MoyBuzKimKni93a,KuaCheCheGe97a} we will restrict to the case in which initially the atom is in its excited state $|+\rangle$ and the electromagnetic field is in a coherent state $|\alpha\rangle=\sum_{n=0}^{\infty}Q_{n}|n\rangle$, with $Q_{n}\equiv\exp(-|\alpha|^{2}/2)\frac{\alpha^{n}}{\sqrt{n!}}$. The unitary evolution ($\gamma=0$) provides the following expressions for these quantities:

\begin{subequations}
\begin{eqnarray}
W(t)&=&\sum_{n=0}^{\infty}|Q_{n}|^{2}\cos\left[2\lambda t\sqrt{\frac{(n+m)!}{n!}}\right]\\
P_{n}(t)&=&|Q_{n}|^2\cos^{2}\left[\lambda t\sqrt{\frac{(n+m)!}{n!}}\right]+\nonumber\\
&+&|Q_{n-m}|^{2}\sin^{2}\left[\lambda t\sqrt{\frac{(n+m)!}{n!}}\right]
\end{eqnarray}
\end{subequations}

The objective is to calculate these same quantities when the phase damping term is present in \eqref{Master}, i.e. when $\gamma\neq 0$. We will make use of the result proved in the previous section and note that the equation \eqref{Master} can be obtained by adding a stochastic term to the original evolution operator and then performing the stochastic average. In our case, $V=\gamma^{1/2}H$, which obviously commutes with the Hamiltonian, thus we are in the first case. The original evolution operator is promoted to 

\begin{eqnarray}
U(t)=\exp(-itH)&\to& U_{st}(t)=\exp(-itH-i\gamma^{1/2}\mathcal{B}_{t}H)\nonumber\\
&=&\exp(-i(t+\gamma^{1/2}\mathcal{B}_{t})H)
\end{eqnarray}

Equivalently we may think that it is $t$ which is promoted $t\to t+\gamma^{1/2}\mathcal{B}_{t}$. Thus to arrive at the desired ``phase-damped'' expressions $W^{pd}(t)$ and $P_{n}^{pd}(t)$ all we have to do is to add a stochastic term to the time variable $t$ and then perform the average:

\begin{subequations}
\begin{eqnarray}
W^{pd}(t)&=&\mathbb{E}\left[\sum_{n=0}^{\infty}|Q_{n}|^{2}\cos\left[2\lambda(t+\gamma^{1/2}\mathcal{B}_{t})\sqrt{\frac{(n+m)!}{n!}}\right]\right]\nonumber\\
&&\\
P_{n}^{pd}(t)&=&\mathbb{E}\left[|Q_{n}|^2\cos^{2}\left[\lambda(t+\gamma^{1/2}\mathcal{B}_{t})\sqrt{\frac{(n+m)!}{n!}}\right]\right.+\nonumber\\
&+&\left.|Q_{n-m}|^{2}\sin^{2}\left[\lambda(t+\gamma^{1/2}\mathcal{B}_{t})\sqrt{\frac{(n+m)!}{n!}}\right]\right]\nonumber\\
\end{eqnarray}
\end{subequations}

Using the linearity property of the expectation value and recalling the moments of the standard real brownian motion (cf. above and \cite{Oksendal98a}), the previous calculations can be carried over elementarily using (see appendix \ref{Appendix}):

\begin{eqnarray}
\mathbb{E}\cos\big[2\lambda(t+\gamma^{1/2}\mathcal{B}_{t})\sqrt{\frac{(n+m)!}{n!}}\big]&=&e^{-2\gamma\lambda^{2}t\frac{(n+m)!}{n!}}\times\nonumber\\
&\times&\cos\left[2\lambda t\sqrt{\frac{(n+m)!}{n!}}\right]\nonumber\\
&&
\end{eqnarray}

Hence

\begin{subequations}
\begin{eqnarray}
W^{pd}(t)&=&\sum_{n=0}^{\infty}|Q_{n}|^{2}e^{-2\gamma\lambda^{2}t\frac{(n+m)!}{n!}}\cos\left[2\lambda t\sqrt{\frac{(n+m)!}{n!}}\right]\nonumber\\
&&\\
P_{n}^{pd}(t)&=&\frac{1}{2}|Q_{n}|^{2}\left\{1+e^{-2\gamma\lambda^{2}t\frac{(n+m)!}{n!}}\right\}\cos\left[2\lambda t\sqrt{\frac{(n+m)!}{n!}}\right]+\nonumber\\
&+&\frac{1}{2}|Q_{n-m}|^{2}\left\{1-e^{-2\gamma\lambda^{2}t\frac{(n+m)!}{n!}}\right\}\cos\left[2\lambda t\sqrt{\frac{(n+m)!}{n!}}\right]\nonumber\\
\end{eqnarray}
\end{subequations}

\noindent which exactly coincides with equations (41) and (43) in \cite{KuaCheCheGe97a} and eqs. (3.26) and (3.27) in \cite{MoyBuzKimKni93a} for $m=1$. We encourage the reader to compare this method with those used in \cite{KuaCheCheGe97a, MoyBuzKimKni93a}.\\
Obviously this formalism can also be used to solve the equation \eqref{Master} with any arbitrary Hamiltonian provided we already know the solution when $\gamma=0$.

\subsection{NonMarkovian Evolution}
\label{NonMarkEvol}
A second consequence of the formalism depicted above is its immediate generalization to nonMarkovian situations. The result in section \ref{LindAsRand} can be readily generalized to the following: \emph{Any Lindbladian master equation, whether Markovian or nonMarkovian, but with selfadjoint Lindblad operators can be obtained as the stochastic average of a random unitary evolution}. The generalization stems out from the single fact that whereas in the Markovian regime we necessarily have to add a stochastic term of the form $\mathcal{B}_{t}V=V\int_{0}^{t}d\mathcal{B}_{t}$, in the nonMarkovian case this restriction drops out and then we may add a term like $V\int_{0}^{t}v(s)d\mathcal{B}_{s}$, where $v(s)$ is an arbitrary real-valued function which encodes e.g. the time response of the environment to the system evolution \footnote{It can be argued that more generality is gained if instead of being a real-valued function, $v(t)$ is a stochastic process. In this case, since we are interested in physical properties which are obtained after averaging, this does not suppose any actual gain.}. Under these circumstances, the previous procedure (for simplicity's sake we will only care about the commuting case; the noncommuting case is similar) drives us to

\begin{equation}
\rho(t)=\exp(t\mathcal{L})\mathbb{E}[\exp\left(-i\mathcal{C}_{V}\int_{0}^{t}v(s)d\mathcal{B}_{s}\right)][\rho(0)] 
\end{equation}

Now developing the exponential again into a power series, calculating the expectation value of each term with some elementary Ito calculus and resumming the series, one arrives at

\begin{equation}
\frac{d\rho(t)}{dt}=-i[H,\rho(t)]-\frac{\gamma(t)}{2}[V,[V,\rho(t)]]
\end{equation}

\noindent where $\gamma(t)=\int_{0}^{t}v^{2}(s)ds$. This is clearly a Lindbladian nonMarkovian master equation. The extension to more than one Lindblad operator is again trivial. This result casts some light into the origin of the Lindbladian structure of master equations with selfadjoint Lindblad operators, independently of their Markovian or nonMarkovian character, something beyond reach of the original axiomatic approach of \cite{Lin76a, GorKosSud76a}.\\
In this sense the result proven here generalizes previous derivation of Lindblad evolution using stochastic calculus \cite{Adl00a,Par92a} by dropping out the semigroup condition. Note that this generalization allows us to conclude that since $\gamma(t)=\int_{0}^{t}v^{2}(s)ds$ the decoherence process is irreversible, i.e. no coherence can be recovered within the domain of validity of the phase-damping ME as an evolution equation for the quantum system.\\
The time dependence of the decoherence factor also suggests a classification of different kinds of environments depending on the rate at which the environment decoheres the system (cf. \cite{SalSan02d}). It remains open what the physical conditions should be to have the different decoherence factors. 




\subsection{Models of intrinsic decoherence}
\label{IntrDecoh}

A third advantage appears as a natural generalization of intrinsic decoherence models already present in the literature \cite{Mil91a,BonOliTomVit00a}. These two models propose an intrinsic mechanism of decoherence based on the random nature of time evolution (we will not enter into the discussion of the physical justification of this hypothesis --see original references for discussion, we will only show how they can be generalized), which basically drives us to the evolution equation \eqref{Master}. The starting hypotheses (apart from the random nature of time evolution and the usual representation of a quantum system by a density operator) are a specific probability distribution \cite{Mil91a} or a semigroup condition (Markovianity) \cite{BonOliTomVit00a} for the time evolution. As a result we obtain a \emph{nondissipative Markovian} master equation in both cases.\\
The formalism presented here dispenses with any of these specific conditions, something which allows us to obtain more general master equations, i.e. both Markovian or nonMarkovian and dissipative or nondissipative.\\
The result comes from the combination of Ito calculus and the spectral representation theorem for unitary operators \cite{SalSan02a}. Let $U(t)$ be an unitary evolution operator. By means of the spectral decomposition theorem \cite{Kreyszig78a} it can be written as

\begin{equation}\label{UnitOp}
U(t)=\int_{-\pi}^{\pi}e^{-i\theta t}dE_{\theta}
\end{equation}

\noindent where $E_{\theta}$ denotes the spectral measure of the evolution operator. Now we perform the stochastic promotion as before by substituting

\begin{equation}
\theta t\to\chi_{t}(\theta)
\end{equation}

where $\chi_{t}(\theta)$ is real stochastic process. Then (see \cite{SalSan02a} for details)
\begin{enumerate}
\item The Markovian nondissipative master equation appearing in \cite{Mil91a} and \cite{BonOliTomVit00a} is obtained if $\chi_{t}(\theta)=\theta t+\gamma^{1/2}\mathcal{B}_{t}$. But note that now the Lindblad operators are not fixed by any initial assumption. If e.g. $\chi_{t}(\theta)=\theta t+\int_{0}^{t}\sigma(s;\theta)d\mathcal{B}_{s}(\theta)$ with correlation function $d\mathcal{B}_{t}(\theta)d\mathcal{B}_{t}(\theta')=e^{-\tau^{2}(\theta-\theta')^{2}}dt$ we arrive at a Lindblad equation

\begin{eqnarray}
\frac{d\rho(t)}{dt}&=&-i[H,\rho(t)]-\nonumber\\
&-&\frac{\gamma}{2}\left(H^{2}\rho(t)+\rho(t)H^{2}-2He^{-\tau^2\mathcal{C}_{H}^{2}}[\rho(t)]H\right)\nonumber\\
\end{eqnarray}

\noindent which is clearly different from the phase-damping master equation \eqref{Master}. Thus even restricting ourselves to the same range of assumptions (Markovianity and nondissipation) we can obtain more master equations.

\item A nonMarkovian nondissipative master equation like e.g. 

\begin{equation}
\frac{d\rho(t)}{dt}=-i[H,\rho(t)]-\frac{\lambda(t)}{2}[H,[H,\rho(t)]]
\end{equation}

\noindent is obtained if $\chi_{t}(\theta)=\theta t+\int_{0}^{t}\sigma(s)d\mathcal{B}_{s}$ and $\lambda(t)\equiv\int_{0}^{t}\sigma^{2}(s)ds$.

\item More general equations can be readily obtained by appropiately combining the correlation properties of $\mathcal{B}_{t}(\theta)$ and the time dependency of $\sigma(t;\theta)$.

\end{enumerate}

This allows these models to be used to explain a wider range of phenomena than that originally considered.

\section{Rabi Oscillations Decay in Cavity QED and Ion Traps Experiments}
\label{RabiQED}

In previous sections we have developed a method to adequately modify the original evolution operator of a quantum system to finally arrive at a Lindbladian master equation. Now we find legitimate to proceed the other way around, i.e. what physical predictions are derived from the assumption that a parameter in the evolution operator of a quantum system is random? To be concrete we will focus upon two different physical systems, namely a Rydberg atom in an optical cavity and a linear rf (Paul) ion trap. We will confront the previous hypothesis with experimental results.

\subsection{The Jaynes-Cummings Model Revisited}
\label{StoJCM}
The JCM model describes the interaction between an atom and the electromagnetic field under very special conditions \cite{JayCum63a, ShoKni93a}. Different generalizations have been proposed to take the model closer to experimental reality keeping its solvability. Among these one can find the inclusion of dissipation (often modelled by coupling the field oscillator to a reservoir of external modes) and/or damping (as a consequence of spontaneous emission), multi-atom, multi-level atom, generalized-interaction and multiple-mode generalizations (see \cite{ShoKni93a} for references).\\
Here we want to introduce a novel proposal, which states that JCM predictions can be rendered more realistic by noticing that the coupling constant $\lambda$ between the atom and the field mode should have a stochastic part which contains part of the effects of the approximations assumed in constructing the original model. Since these effects are not under control, to make physical predictions we must average on the introduced random parameters. To illustrate the idea let us consider the original JCM with Hamiltonian $H=\omega a^{\dagger}a+\omega_{0}S_{z}+\lambda(S_{-}a^{\dagger}+S_{+}a)$, where $\omega$ denotes the frequency of the field mode, $a$ and $a^{\dagger}$ their corresponding creation and destruction operators, $\omega_{0}$ the frequency difference between the two energy atomic levels, $S_{z}$ the atomic population operator, $\lambda$ the atom-field coupling constant and $S_{\pm}$ the energy raising/lowering atomic operators. We claim that the evolution stemming out from $H$ should be modified by inserting a random part $\chi_{t}$, where $\chi_t$ is a real stochastic process which contains the departure from the original ideal situation. The connection with the previous formalism is established by noticing that the evolution operator (in interaction picture) must then be:

\begin{eqnarray}
U_{I}(t)&=&\mathcal{T}\exp\left[-i\int_{0}^{t}(\lambda+\chi_{s}) H_{I}^{int}(s)ds\right]=\nonumber\\
&=&\exp\left[-i(\lambda t+\int_{0}^{t}\chi_{s}ds)H^{int}\right]
\end{eqnarray}

\noindent where $H^{int}=(S_{-}a^{\dagger}+S_{+}a)$ is the interaction Hamiltonian and $H^{int}_{I}(t)=U_{0}^{\dagger}(t)H^{int}U_{0}(t)$ is the interaction Hamiltonian in interaction picture (for simplicity's shake exact resonance has been assumed $\omega=\omega_{0}$). Now the expression $\bar{\lambda}_{t}\equiv\lambda t+\int_{0}^{t}\chi_{s}ds$ is a real stochastic process itself which can always be expressed in the form \cite{Oksendal98a} 

\begin{equation}
\bar{\lambda}_{t}=\mathbb{E}[\bar{\lambda}_{t}]+\int_{0}^{t}v_{s}d\mathcal{B}_{s}
\end{equation}

\noindent where $v_{t}$ is a real stochastic process uniquely determined by $\bar{\lambda}_{t}$. Then the density operator in interaction picture will then be given by

\begin{equation}\label{GenDensOp}
\rho_{I}(t)=\mathbb{E}\left[\exp\left(-i\int_{0}^{t}\left(\mathbb{E}[\bar{\lambda}_{s}]+v_{s}d\mathcal{B}_{s}\right)\mathcal{C}_{H^{int}_{I}(s)}\right)ds\right][\rho(0)]
\end{equation}

Instead of giving the general form of the expectation value in \eqref{GenDensOp} (which will be difficult to obtain in full generality), we will propose some physical choices based on $\chi_{t}$. If $\chi_{t}=\gamma^{1/2}\xi_{t}$, i.e. the original deterministic evolution is  randomly perturbed by a white noise coming from a stochastic perturbation of the coupling constant, then $\bar{\lambda}_{t}=\lambda t+\gamma^{1/2}\mathcal{B}_{t}$ and \eqref{GenDensOp} reduces to

\begin{eqnarray}
\rho_{I}(t)&=&\exp\left(-i\lambda t\mathcal{C}_{H_{I}}\right)\mathbb{E}[\exp\left(-i\gamma^{1/2}\mathcal{B}_{t}\mathcal{C}_{H_{I}}\right)[\rho(0)]\nonumber\\
&=&\mathbb{E}[\exp\left(-i\left(\lambda t+\gamma^{1/2}\mathcal{B}_{t}\right)\mathcal{C}_{H_{I}}\right)][\rho(0)]
\end{eqnarray}

\noindent which yields a density operator in Schr\"{o}dinger picture given by

\begin{equation}
\rho(t)=\mathbb{E}[\exp\left(-it\mathcal{C}_{H_{0}}-i\left(\lambda t+\gamma^{1/2}\mathcal{B}_{t}\right)\mathcal{C}_{H_{I}}\right)][\rho(0)]
\end{equation}

This relationship means that to obtain the physical predictions of this proposal, all one has to do is to make the sustitution $\lambda t\to\lambda t+\gamma^{1/2}\mathcal{B}_{t}$ in the original expressions and calculate the expectation value.\\
Note that this proposal allows us to embrace nonMarkovian (though Lindbladian) situations with little extra effort, e.g. by claiming that the random perturbation is time-dependent $\chi_{t}=\gamma^{1/2}(t)\xi_{t}$. More general options are also possible. The particular choice for $\chi_{t}$ relies upon the specific system under study. Notice also that the generalization proposed here is compatible with the ones quoted above, i.e. one may combine both type of generalizations.

\subsection{Decay in an Optical Cavity}
\label{QED}
\begin{center}
\begin{figure}[ht]
\epsfig{file=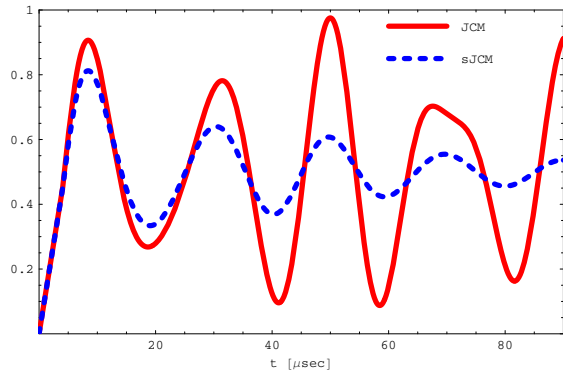,width=8.0cm} \caption{\label{QEDRabi}$P_{eg}(t)$ for the original and the stochastic JCM's with parameters $\lambda=50\pi kHz$, $\gamma=1/2\pi$ and $|\alpha|^{2}=\bar{n}_{photons}=0.4$. See \cite{BruSchmMaaDreHagRaiHar96a}.}
\end{figure}
\end{center}

Let us consider the situation depicted in \cite{BruSchmMaaDreHagRaiHar96a}, which appears as the first direct (in time domain) experimental evidence of field quantization. The system consists of a Rydberg atom in a high-Q optical cavity with the atom initially excited and the electromagnetic field in a coherent state $|\alpha\rangle$. This system is accurately described using the JCM, thus Rabi oscillations are expected and concordantly experimentally measured. The theoretical prediction for the probability $P_{eg}(t)$ to find the atom at a time $t$ in its ground state is $P_{eg}(t)=\sum_{n=0}^{\infty}|Q_{n}|^{2}\sin^{2}\left(\lambda t\sqrt{n+1}\right)$.  However an exponential damping in these oscillations are detected (see \cite{BruSchmMaaDreHagRaiHar96a} for details). The stochastic JCM accounts for this damping (see fig. \ref{QEDRabi}) assuming $\chi_{t}=\gamma^{1/2}\xi_{t}$ which produces 

\begin{equation}
P_{eg}^{sJCM}(t)=\frac{1}{2}\left(1-\sum_{n=0}^{\infty}|Q_{n}|^{2}e^{-2(n+1)\gamma\lambda t}\cos\left(2\lambda t \sqrt{n+1}\right)\right)
\end{equation}

The physical interpretation under this assumption is rather clear: the ideal coupling assumed in the original JCM does not hold any longer and departures from this ideality should be considered. Darks counts and decoherence caused by collisions with background gas have been considered as candidates to explain the damping behaviour\footnote{Notice that in the experimental conditions achieved, spontaneous emission cannot play any role to this extent. See \cite{BruSchmMaaDreHagRaiHar96a}.} and even more radical proposals appear in the literature \cite{BonOliTomVit00a}. Except for the latter which reveals an original intrinsic process, all of them resort to external agents. Here note that we do not need to do so, since the departure from ideal atom-field mode coupling can be justified within the domain of the JCM assumptions themselves, i.e. the assumption of coupling between a unique field mode and two levels of the atom can be relaxed by adding in a natural way a random background in this coupling.  

This decay is also present in the case of arbitrary initial conditions. If $P_{eg}(t)=\sum_{n=0}^{\infty}P_{n}\sin^{2}\left(\lambda t\sqrt{n+1}\right)$ is the orthodox prediction, then the previous recipe drives us to 

\begin{equation}
P_{eg}^{sJCM}(t)=\frac{1}{2}\left(1-\sum_{n=0}^{\infty}P_{n}e^{-2(n+1)\gamma\lambda t}\cos\left(2\lambda t \sqrt{n+1}\right)\right)
\end{equation}

\noindent where $P_{n}$ depends on the actual initial conditions of both the atom and field mode. Note that not only can any actual exponential decay (with arbitrary time dependency) be obtained with the substitution $\lambda t\to\lambda t+\int_{0}^{t}\gamma^{1/2}(s)d\mathcal{B}_{s}$, but also possible changes in the argument of the cosine function could be accounted for by making $\lambda t\to\omega(t)+\int_{0}^{t}\gamma^{1/2}(s)d\mathcal{B}_{s}$. See appendix \ref{Appendix} for details.\\
In this way we have obtained a decohering system (oscillations coming from quantum superpositions are progressively supressed) without necessarily resorting to the action of the environment and keeping quantum principles untouched (see discussion in section \ref{Discuss} later on). This opens new possibilities to discuss possible sources of decoherence. 


\subsection{Decay in an Ion Trap}
\label{IonDecay}
The previous experimental supression of quantum coherence has also been detected in a linear Paul ion trap \cite{MeeMonKinItaWin96a,WinMonItaLeiKinMee98a}. The physical situation is formally similar to that of the Rydberg atom coupled to a field mode: the laser field is operated upon the trap in such a way that it can be considered that only two internal energy levels of the ions are coupled to the center-of-mass (COM) mode of the set of ions (see \cite{WinMonItaLeiKinMee98a}). In the dipole and rotating-wave approximations, the interaction Hamiltonian (in interaction picture) is 

\begin{equation}\label{IntHamilIntPic}
H_{I}^{int}(t)=\Omega S_{+}\exp\left(i\left[\eta (a e^{-i\omega_{z}t}+a^{\dagger} e^{i\omega_{z}t})-\delta t\right]\right)+\textrm{h.c.} 
\end{equation}

\noindent where $\eta\equiv kz_{0}$ is the Lamb-Dicke parameter ($k=2\pi/\lambda_{laser}$ and $z_{0}\equiv(\langle 0|z|0\rangle^{1/2}$), $S_{\pm}$ denote the raising/lowering operators for the internal levels, $a$ and $a^{\dagger}$ denote the destruction and creation operators for the COM mode, $\omega_{z}$ denotes the frequency of the harmonic trap for the COM, $\delta=\omega-\omega_{0}$ with $\omega$ the frequency of the laser mode and $\omega_{0}$ denotes the difference between the two internal energy levels of the ions.\\
The statevector can then be written as 

\begin{equation}
|\Psi(t)\rangle=\sum_{m_{z}=\pm}\sum_{n=0}^{\infty}C_{m_{z},n}(t)|m_{z}\rangle\otimes|n\rangle
\end{equation}

\noindent where $|m_{z}\rangle$ and $|n\rangle$ denote the (time-independent) internal and motional eigenstates. In the conditions of interest, i.e. in resonant transitions ($\delta=\omega_{z}(n'-n)$ with $n,n'$ integers), the coefficients $C_{m_{z},n}(t)$ satisfy the equations \cite{WinMonItaLeiKinMee98a}

\begin{eqnarray}
\dot{C}_{+,n'}&=&-i^{1+|n-n'|}\Omega_{n,n'}C_{-,n}\\
\dot{C}_{-,n'}&=&-i^{1-|n-n'|}\Omega_{n',n}C_{+,n'}
\end{eqnarray}

\noindent where $\Omega_{n,n'}$ is given by $\Omega_{n,n'}\equiv\Omega|\langle n'|e^{ikz}|n\rangle|$ ($k=2\pi/\lambda_{\textrm{laser}}$ and $z=z_{0}(a+a^{\dagger})$ is the operator for the COM motion). From these one can predict the well-known Rabi oscillations of the system. For concreteness' shake let us focus upon the first blue sideband case, i.e. when $n'=n+1$. If the trap is prepared in the initial state $|\Psi(0)\rangle=|-\rangle\otimes|n\rangle\equiv|-,n\rangle$, then the probability of finding a single ion in the $|-\rangle$ state at time $t$ is $P_{-}(t)=\cos^{2}(\Omega_{n,n+1}t)$. However experimentally an exponential decay is obtained. As before, one can argue that ideality should be restricted and both a substitution $\Omega_{n,n+1}t\to\Omega_{n,n+1}t+\int_{0}^{t}\gamma^{1/2}(s)d\mathcal{B}_{s}$ and the corresponding averaging should be performed on $P_{-}(t)$. This would drives us to the relation 

\begin{equation}
P^{rand}_{-}(t)=\frac{1}{2}\left(1+e^{-2\lambda(t)}\cos(2\Omega_{n,n+1}t)\right)
\end{equation}

\noindent where $\lambda(t)\equiv\int_{0}^{t}\gamma(s)ds$. This way the exponential decay would have been obtained. Physically this recipe can be justified by taking into account intensity fluctuations in the laser field (see \cite{SchneiMil97a} --notice that some generality is gained with respect to this work).\\
However experimental data for the COM initially in an arbitrary state and the ion in the ground state $|-\rangle$ are better fit by 

\begin{equation}
P_{-}^{exp}(t)=\frac{1}{2}\left[1+\sum_{n=0}^{\infty}P_{n}e^{-\gamma_{n}t}\cos(2\Omega_{n,n+1}t)\right]
\end{equation}  

\noindent where $P_{n}$ is an $n$-dependent quantity which relies upon the initial conditions of the ion's motion and $\gamma_{n}=\gamma_{0}(n+1)^{0.7}$ is a phenomenologically decoherence rate \cite{MeeMonKinItaWin96a}. The peculiar exponent $0.7$ in $\gamma_{n}$ renders the previous physical explanation insufficient. More involved schemes to account for this exponent can already be found in the literature \cite{MurKni98a,BonOliTomVit00a}. Here we propose a new one based on the previously introduced random evolution schemes.\\
The main problem attains the peculiar $n$-dependency of the argument of the exponential decaying function. In the mathematical realm the necessary flexibility comes from a combination of stochastic calculus and the spectral theorem for the evolution operator \cite{SalSan02a} and in the physical one from realizing that not all energy levels of the COM mode can be equally affected by a stochastic perturbation. This idea, in a different context in which the trap is coupled to a boson reservoir to account for the detected decoherence, has already been paid attention \cite{MurKni98a}.\\
Let's start by considering the spectral decomposition \cite{Kreyszig78a}  of the evolution operator generated by the Hamiltonian \eqref{IntHamilIntPic} when the laser is tuned to the first blue sideband, i.e. when the Hamiltonian is given by 

\begin{equation}\label{FirstRedSideBand}
H_{I}^{int}=\eta\Omega[S_{+}a^{\dagger}+S_{-}a]
\end{equation}

Then the evolution operator will be decomposed as follows:

\begin{equation}\label{EvolOpIntPic}
U^{int}_{I}(t)=\sum_{\substack{n=0\\m_{z}=\pm}}^{\infty}e^{-ie_{n}^{m_{z}}t}P_{n}^{m_{z}}
\end{equation} 

\noindent where $e_{n}^{\pm}\equiv \pm\eta\Omega\sqrt{n}$, $|e_{n}^{\pm}\rangle\equiv 1/\sqrt{2}\left(|-,n-1\rangle\pm|+,n\rangle\right)$ ($n\geq 1$) and $|e_ {0}^{+}\rangle=|+,0\rangle$ denote the eigenvalues and eigenstates of \eqref{FirstRedSideBand} respectively and $P_{n}^{m_{z}}=|e_{n}^{m_{z}}\rangle\langle e_{n}^{m_{z}}|$ is the projector-valued measure associated to \eqref{FirstRedSideBand}. The stochastic promotion is performed by substituting $e_{n}^{m_{z}}t\to e_{n}^{m_{z}}t+\int_{0}^{t}v_{n}^{m_{z}}(s)d\mathcal{B}_{n}^{m_{z}}(s)$ in \eqref{EvolOpIntPic} and calculating $\rho_{int}(t)=\mathbb{E}\left[U^{int}_{I,rand}(t)\rho(0)(U^{int}_{I,rand}(t))^{\dagger}\right]$. Here note that the different energy levels are perturbed in a distinct fashion determined both by the deterministic functions $v_{n}^{m_{z}}(t)$ and the standard real Brownian motions $\mathcal{B}_{n}^{m_{z}}(t)$. The latter show  correlation properties expressed by the functions $g_{n,m}^{m_{z},m_{z}^{'}}(t)$:

\begin{equation}
d\mathcal{B}_{n}^{m_{z}}(t)d\mathcal{B}_{m}^{m_{z}^{'}}(t)=g_{n,m}^{m_{z},m_{z}^{'}}(t)dt
\end{equation}

The density operator in interaction picture will then be given by

\begin{equation}\label{PreCal}
\rho_{int}(t)=\sum_{\substack{n,m=0\\m_{z},m_{z}^{'}=\pm}}^{\infty}e^{-i(e_{n}^{m_{z}}-e_{m}^{m_{z}^{'}})t}\mathbb{E}\left[e^{-i\left(\int_{0}^{t}v_{n}^{m_{z}}(s)d\mathcal{B}_{n}^{m_{z}}(s)-\int_{0}^{t}v_{m}^{m_{z}^{'}}(s)d\mathcal{B}_{m}^{m_{z}^{'}}(s)\right)}\right]P_{n}^{m_{z}}\rho(0)P_{m}^{m_{z}^{'}}
\end{equation}

\begin{center}
\begin{figure}[htb]
\epsfig{file=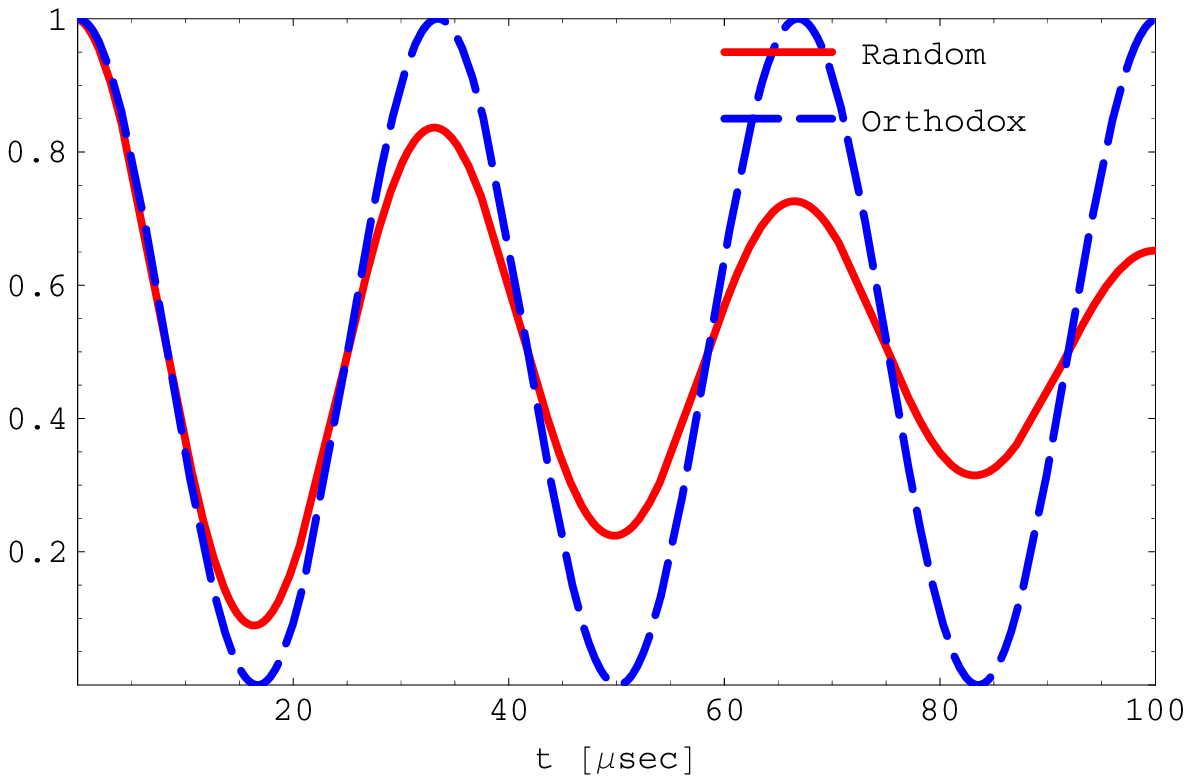,width=7.5cm} \caption{\label{FockGraph}$P_{-}(t)$ for the orthodox and the stochastic formalisms with $\rho(0)=|-,n=0\rangle\langle-,n=0|$ and parameters $\Omega= 470kHz$, $\gamma_{0}=11.9kHz$ and $\eta=0.202$. See \cite{MeeMonKinItaWin96a}.}
\end{figure}
\end{center}

The expectation value in \eqref{PreCal} can be calculated with the same techniques as before (cf. also  appendix \ref{Appendix}) and drives us to:

\begin{equation}\label{DensOpWithDecoh}
\rho_{int}(t)=\sum_{\substack{n,m=0\\m_{z},m_{z}^{'}=\pm}}^{\infty}e^{-i(e_{n}^{m_{z}}-e_{m}^{m_{z}^{'}})t-\frac{\Lambda_{n,m}^{m_{z},m_{z}^{'}}(t)}{2}}P_{n}^{m_{z}}\rho(0)P_{m}^{m_{z}^{'}}
\end{equation}

\noindent where $\Lambda_{n,m}^{m_{z},m_{z}^{'}}(t)=\int_{0}^{t}\lambda_{n,m}^{m_{z},m_{z}^{'}}(s)ds$ with $\lambda_{n,m}^{m_{z},m_{z}^{'}}(t)=(v_{n}^{m_{z}}(t))^{2}g_{n,n}^{m_{z},m_{z}}(t)+(v_{m}^{m_{z}^{'}}(t))^{2}g_{m,m}^{m_{z}^{'},m_{z}^{'}}(t)-2v_{n}^{m_{z}}(t)v_{m}^{m_{z}^{'}}(t)g_{n,m}^{m_{z},m_{z}^{'}}(t)$.

\begin{center}
\begin{figure}[htb]
\epsfig{file=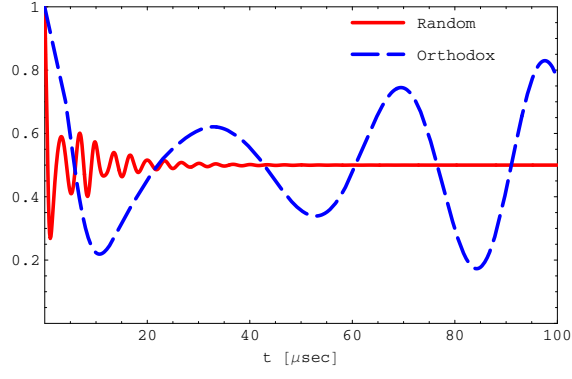, width=7.5cm} \caption{\label{ThermalGraph}$P_{-}(t)$ for the orthodox and the stochastic formalisms with $\rho(0)=|-\rangle\langle-|\otimes\rho_{T}$ and parameters $\Omega= 470kHz$, $\gamma_{0}=11.9kHz$, $\eta=0.202$ and $\bar{n}_{photons}= 1.5$. See \cite{MeeMonKinItaWin96a}.}
\end{figure}
\end{center}

The expression \eqref{DensOpWithDecoh} already contains the necessary ingredients to arrive at the detected behaviour, since if the ion trap is initially set in a Fock state for the COM mode and the ground state for the internal levels, i.e. $\rho(0)=|-,n\rangle\langle -,n|$, then the probability $P_{-}^{rand}(t)$ in this scheme is

\begin{widetext}
\begin{eqnarray}\label{ProbRand}
P_{-}^{rand}(t)&=&\frac{1}{4}\left[\exp\left\{-\frac{\Lambda_{n+1,n+1}^{+,+}(t)}{2}\right\}+\exp\left\{-\frac{\Lambda_{n+1,n+1}^{-,-}(t)}{2}\right\}\right.\nonumber\\
&+&\left.\exp\left\{-i(e_{n+1}^{+}-e_{n+1}^{-})t-\frac{\Lambda_{n+1,n+1}^{+,-}(t)}{2}\right\}+\exp\left\{-i(e_{n+1}^{-}-e_{n+1}^{+})t-\frac{\Lambda_{n+1,n+1}^{-,+}(t)}{2}\right\}\right]\nonumber\\
\end{eqnarray}
\end{widetext}

Before making physical assumptions let us notice that since the Brownian motions are standard, $g_{n,n}^{(\pm,\pm)}(t)=1$ for all $n$ (the brackets mean that both superscripts must be equal) and thus $\Lambda_{n,n}^{(\pm,\pm)}(t)=0$ for all $n$. Now we pose the most important physical hypothesis, namely the stochastic perturbation depends exclusively upon the energy level of the COM mode (at least up to the order of detection we are nowadays capable). As a first consequence we then can claim that $g_{n,n}^{+,-}(t)=g_{n,n}^{-,+}(t)$ for all $n$, and then $\lambda_{n,n}^{(\pm,\pm)}(t)=0$ and  $\Lambda_{n,n}^{(\pm,\pm)}(t)=0$ and also $\Lambda_{n,n}^{+,-}(t)=\Lambda_{n,n}^{-,+}(t)$, hence \eqref{ProbRand} reduces to 

\begin{eqnarray}
P_{-}^{rand}(t)&=&\frac{1}{2}\left[1+e^{-\frac{\Lambda_{n+1,n+1}^{+,-}(t)}{2}}\cos((e_{n+1}^{+}-e_{n+1}^{-})t)\right]=\nonumber\\
&=&\frac{1}{2}\left[1+e^{-\frac{\Lambda_{n+1,n+1}^{+,-}(t)}{2}}\cos(2\eta\Omega t\sqrt{n+1})\right]
\end{eqnarray}

\begin{center}
\begin{figure}[htb]
\epsfig{file=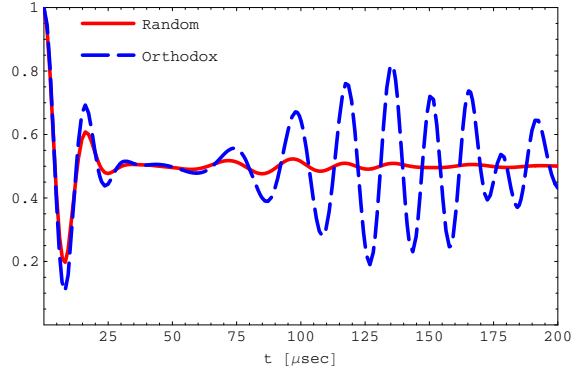, width=7.5cm}\caption{\label{CoherGraph}$P_{-}(t)$ for the orthodox and the stochastic formalisms with $\rho(0)=\sum_{n,m}Q_{n}Q_{m}^{*}|-,n\rangle\langle-,m|$ and parameters $\Omega= 470kHz$, $\gamma_{0}=11.9kHz$, $\eta=0.202$ and $\bar{n}= 1.5$. See \cite{MeeMonKinItaWin96a}.}
\end{figure}
\end{center}

Second since for fixed $n$ the internal levels are equally affected, we can also write $g_{n,n}^{+,-}(t)=g_{n,n}^{+,+}(t)$ for each $n$. Finally instead of discussing upon absolute energy values, it is physically more reasonable to talk about energy differences  and we propose that the stochastic perturbations be introduced in such a way as to have

\begin{equation}
e_{n}^{+}-e_{n}^{-}\to e_{n}^{+}-e_{n}^{-}+\Gamma^{1/2}n^{d}\xi_{t}
\end{equation} 

\noindent where $d$ is an arbitrary exponent, $\Gamma$ a constant and where we have assumed for simplicity that the random perturbation is a white noise. Note however that it is possible to use more general expressions. Notice the different behaviour of the added term for each distinct subspace of constant COM energy in agreement with the physical hypothesis assumed above. Under these hypotheses $v_{n}^{+}(t)-v_{n}^{-}(t)=\Gamma^{1/2}n^{d}$ and after elementary calculations \eqref{ProbRand} finally reduces to

\begin{equation}
P_{-}^{rand}(t)=\frac{1}{2}\left(1+e^{-\frac{\Gamma t}{2}(n+1)^{2d}}\cos\left(2\eta\Omega t\sqrt{n+1}\right)\right)
\end{equation}

This expression shows a clear resemblance to $P_{-}^{exp}(t)$ written above. We believe that both $\Gamma$ and $d$ depends sensitively upon the particular physical system under study.\\
For completeness we also include the expression for $P_{-}^{rand}(t)$ when the COM mode has an initial state with diagonal density-matrix elements $P_{n}$:

\begin{equation}
P_{-}^{rand}(t)=\frac{1}{2}\left(1+\sum_{n=0}^{\infty}P_{n}e^{-\frac{\Gamma t}{2}(n+1)^{2d}}\cos\left(2\eta\Omega t\sqrt{n+1}\right)\right)
\end{equation}

This expression has the same structure as the experimental ones shown in \cite{MeeMonKinItaWin96a}. The whole scheme can obviously be applied to the carrier, first red sideband and successive excitation too. We include in figs. \ref{FockGraph}, \ref{ThermalGraph} and \ref{CoherGraph} the predictions in the orthodox and the above formalism in the cases when the COM mode is in a Fock state, thermal state and coherent state and the internal state is the ground state.

\section{Discussion}
\label{Discuss}
The use of stochastic methods in Hilbert space is of course not new (cf. 
e.g. \cite{Dio86a,DioLuc94a}). The 
idea of representing open quantum systems by means of stochastic processes 
already appeared in the literature some years ago \cite{Dio85a} and it 
has been widely used in Quantum Optics \cite{PleKni98a} and in the 
Foundations of Quantum Mechanics \cite{Pea99a}. Here we 
pursue the line initiated in \cite{Dio86a} stepping forward by randomizing 
not just the (thus stochastic) state vector of the open quantum system, 
but its evolution operator. We find at least three reasons to do that. 
Firstly when one write a random evolution equation for the state vector 
(thus an Ito stochastic differential equation) an extra term must be 
included, namely the Ito correction. Consider for example the following 
evolution

\begin{equation}\label{ItoState}
d|\psi_{t}\rangle=-iH|\psi_{t}\rangle dt-iA|\psi_{t}\rangle d\mathcal{B}_{t}-\frac{1}{2}A^{2}|\psi_{t}\rangle dt
\end{equation} 

\noindent where the operator $A$ commutes with the hamiltonian $H$ (just 
for simplicity). From a 
physical point of view we find little intuitive the origin of the Ito 
correction term, which however appears in a natural way by applying Ito's 
formula to the evolution operator with the stochastic modification 
$U_{st}(t)=e^{-iHt-iA\mathcal{B}_{t}}$. Secondly the use of random 
evolution operators emphasizes the idea that it is the evolution which is 
random and there is nothing random about the Hamiltonian (and thus the 
energy levels of the system), something which may misleadingly be 
understood from equation \eqref{ItoState}. Finally the use of operators 
rather than just state vectors opens the possibility of trying to employ 
group-representation techniques \cite{Bal90a} and thus of rooting the 
random nature of the evolution upon possible stochastic symmetries.\\
This proposal is not intended to solve the so-called 
\emph{macroobjectification} problem (better known as the measurement 
problem) by means of a random dynamical reduction process. Indeed it can be readily shown that 
the state driven by a random evolution operator 
$U_{st}(t)=e^{-iHt-iV\mathcal{B}_{t}}$ is never reduced in clear contrast to these models (see \cite{AdlBroBruHug01a}). For the case of the previous random evolution operator this can be readily proven. Let $U_{st}(t)=e^{-iHt-iA\mathcal{B}_{t}}$ be the random evolution operator of a quantum system ($[A,H]=0$ for simplicity). Then the Ito differential equation for the state vector is \eqref{ItoState}. Now to check whether this evolution produces dynamical state-reduction or not it is sufficient to study the stochastic process (cf. \cite{AdlBroBruHug01a})

\begin{equation}
\tilde{V}_{t}\equiv\frac{\langle\psi_{t}|(A-A_{t})^{2}|\psi_{t}\rangle}{\langle\psi_{t}|\psi_{t}\rangle}
\end{equation}

\noindent where $A_{t}\equiv\frac{\langle\psi_{t}|A|\psi_{t}\rangle}{\langle\psi_{t}|\psi_{t}\rangle}$. Since $d\langle\psi_{t}|\psi_{t}\rangle=0$ and $|\langle\psi_{0}|\psi_{0}\rangle|^{2}=1$, the random evolution is unitary almost surely and then

\begin{equation}
d\tilde{V}_{t}=0\Rightarrow\tilde{V}_{t}=\tilde{V}_{0}
\end{equation}

Thus there is no actual state-vector reduction process around the eigenvectors of $A$. This of course differs radically for the behaviour of $|\psi_{t}\rangle$ in stochastic state vector reduction models, where the evolution equation is typically written as \cite{AdlBroBruHug01a,Pea99a}

\begin{equation}\label{ItoRed}
d|\psi_{t}\rangle=-iH|\psi_{t}\rangle dt-\frac{1}{2}(A-A_{t})^{2}|\psi_{t}\rangle+(A-A_{t})|\psi_{t}\rangle d\mathcal{B}_{t}
\end{equation}

The nonlinear terms appear as a consequence of normalization conditions \cite{SalSan02c}  and play no significant role in the reduction process. Note the singular difference between eqs. \eqref{ItoState} and \eqref{ItoRed}: despite the fact that both of them produce the same master equation (already noted in \cite{DioWis01a}), only the second one ensures a reduction process taking place and this is because of the $i$ factor appearing with the Wiener differential $d\mathcal{B}_{t}$. It is an open question whether there exists a physical process or not introducing this phase factor in the evolution equation for $|\psi_{t}\rangle$.\\
The great utility of stochastic processes to account for the decoherence 
suffered by a quantum system is its versatility to also account for 
possible intrinsic decohering effects. By this we mean not a fundamental 
modification of quantum principles, as e.g. in 
\cite{Mil91a,BonOliTomVit00a,Pea99a} but just the idea 
that when a quantum system is described (e.g. an atom interacting with the 
electromagnetic field) some approximations must necessarily be made to be 
able to analytically handle with it and we claim that some of these 
approximations may hidingly induce decohering effects upon the 
approximated model. In this sense this decoherence can be called 
\emph{intrinsic} 
since no environmental effect is taken into account. Undoubtedly this does 
not deny in any way the possibility of having a system decohered by its 
environment. Notice the special relevance of such a hypothetical effect in 
quantum systems designed to implement quantum-computational and 
quantum-information-processing tasks, since they usually possess certain 
degree of complexity which forces us to seek for adequate approximations to 
describe them. A possible relationship of this intrinsic decoherence with scalability of quantum-informational and quantum computational systems would also have important consequences to find robust mechanisms to process information in a quantum way.\\
Besides possible new physical interpretations supporting the use of 
stochastic processes in Quantum Mechanics, its utility to solve certain ME as
 shown above justifies the search to extend the main result reported here. 
Currently the way to drop the condition of selfadjointness of Lindblad 
operators is under study.\\
Finally a mathematical remark should be made regarding the proof of the 
previous result. One may wonder why the same procedure as in the commuting 
case, i.e. adding a stochastic term $U(t)\to 
e^{-iHt-V\int_{0}^{t}\sigma(s)d\mathcal{B}_{s}}$, is not used in the 
noncommuting case. The reason jointly rests upon the Ito's formula and the 
lack of some derivatives in the case of noncommuting operators. Ito's 
formula \cite{Oksendal98a} basically states \footnote{Differentiablity 
conditions are supposed being enough satisfied as to guarantee the existence of the performed operations. See \cite{Oksendal98a,KarShr91a} for details.} 
that if the real stochastic process $X_{t}$ 
satisfies the Ito SDE 
$dX_{t}=a(t,X_{t})dt+b(t,X_{t})d\mathcal{B}_{t}$, then the real stochastic 
process defined as $Y_{t}=\psi(t,X_{t})$ satisfies the Ito SDE given 
by

\begin{equation}
d\psi(t,X_{t})=\left(a(t,X_{t})\partial_{t}\psi(t,X_{t})+\frac{1}{2}b^{2}(t,X_{t})\partial_{xx}\psi(t,X_{t})\right)dt+b(t,X_{t})\partial_{x}\psi(t,X_{t})d\mathcal{B}_{t}
\end{equation}    

This means that both the first and second partial derivatives of 
$\psi(t,x)$ must exist to be able to apply this formula. If the stochastic 
process is operator-valued as e.g.  $e^{-itH-iX_{t}V}$ with $[H,V]\neq 
0$, then Ito's formula cannot be directly applied since the 
partial derivates of $e^{-itH-ixV}$ cannot be found. This difficulty has 
been circumvented by changing to the Heisenberg picture before 
introducing the stochastic modifications. However it would be desirable to 
have an Ito's formula for operator-valued stochastic processes valid both 
for commuting and noncommuting cases.

\section{Conclusions}

We have proven how any Lindblad evolution with selfadjoint Lindblad operators can be understood as an averaged random evolution operator. The proof included here allows us to extend the previous result to nonMarkovian situations, though keeping the Lindbladian structure of the master equations, and as a result we have provided a straightforward method to solve this kind of master equations. The conjunction of stochastic methods and the spectral representation theorem has also allowed us to generalize intrinsic decoherence models already present in the literature. The mathematical versatility of stochastic methods has also permitted us to propose a generalization to the Jaynes-Cummings model, and compare its predictions with experimental results in cavity QED as well as in ion traps. We have argued that the main physical advantage stems from the possibility of studying decohering effects not necessarily rooted on the environmental action upon the system. A comparison with dynamical collapse models reveals that the random unitary operators used here do not produce any kind of state vector reduction. 

\begin{acknowledgments}
One of us (D.S.)  acknowledges the support of Madrid Education Council under grant BOCAM 20-08-1999. 
\end{acknowledgments}

\appendix

\section{\label{Appendix}Some Useful Identities}
\label{Append}
Though they are elementary we include some useful relations in order to render the text self-contained. First we will show how the moments of arbitrary order $n$ of the stochastic process $\int_{0}^{t}v(s)d\mathcal{B}_{s}$ --$v(t)$ a real function-- are calculated. Let us denote the stochastic process and their $n$th moments respectively  as $X_{t}\equiv\int_{0}^{t}v(s)d\mathcal{B}_{s}$ and $\beta_{n}(t)\equiv\mathbb{E}\left[\int_{0}^{t}v(s)d\mathcal{B}_{s}\right]^{n}$. It is evident that $X_{0}=0$. $X_{t}$ satisfies the Stochastic Differential Equation $dX_{t}=v(t)d\mathcal{B}_{t}$. Applying Ito's formula \cite{Oksendal98a} to $f(X_{t})$ with $f(x)=x^{n}$ and taking expectation values one readily arrives at

\begin{subequations}
\begin{gather}
\beta_{2n}(t)=\frac{(2n)!}{2^n n!}\lambda^{n}(t)\\
\beta_{2n+1}(t)=0
\end{gather}
\end{subequations}

\noindent where $\lambda(t)=\int_{0}^{t}v^{2}(s)ds$.\\
Using the well-known formula $\cos(A+B)=\cos A\cos B-\sin A\sin B$, developing the trigonometric functions into a power series and using the previous result one can immediately arrive at

\begin{equation}\label{RandCos}
\mathbb{E}[\cos(b(t)+\int_{0}^{t}v(s)d\mathcal{B}_{s})]=e^{-\frac{\lambda(t)}{2}}\cos b(t)
\end{equation}

To calculate $\mathbb{E}[\cos^{2}Z_{t}]$, it is convenient to use $\cos^{2}A=\frac{1+\cos 2A}{2}$ and then apply \eqref{RandCos}.\\
Finally theorem 3 of \cite{Lou73a} is quoted with the notation introduced in section \ref{LindAsRand}:

\begin{theor}
If $A$ and $B$ are two fixed noncommuting operators and $\xi$ is a parameter, then

\begin{equation}\label{TheoLouisell}
e^{\xi A}Be^{-\xi A}=e^{\xi\mathcal{C}_{A}}[B]
\end{equation}
\end{theor}

See original reference for the proof.


\end{document}